\documentclass[twocolumn,showpacs,preprintnumbers,amsmath,amssymb]{revtex4}
\usepackage{graphicx}
\usepackage{dcolumn}
\usepackage{bm}

\newcommand{\er}{{\bf e}_{r}}
\newcommand{\ep}{{\bf e}_\phi}
\newcommand{\ez}{{\bf e}_z}
\newcommand{\hh}{{\bf H}}
\newcommand{\ee}{{\bf E}}
\newcommand{\s}{{\bf S}}

\newcommand{\rr}{{\bf r}}

\newcommand{\rmi}{{\rm i}}

\begin{document}

\title{Optical Poynting singularities of propagating and evanescent vector Bessel beams}

\author{Denis V. Novitsky}
\email{dvnovitsky@tut.by}
\affiliation{%
B.I. Stepanov Institute of Physics, National Academy of Sciences of
Belarus, \\ Nezavisimosti~Avenue~68, 220072 Minsk, Belarus.
}%

\author{Andrey V. Novitsky}
\email{andrey.novitsky@tut.by}
\affiliation{%
Belarusian State University, Department of Theoretical Physics, \\
Nezavisimosti~Avenue~4, 220050 Minsk, Belarus.
}%

\begin{abstract}
For propagating and evanescent vector Bessel beams, we study the
singularities of the Poynting vector (Poynting singularities), at
which the energy flux density turns to zero. Poynting singularities
include all the phase singularities and some of polarization ones
(L- and C-points). We reveal the existence conditions and positions
of singularities, which are located at cylindrical surfaces around
the beam axis. We mark the special case of the evanescent Bessel
beam in the form of cylindrical standing wave, that is singular at
any spatial point.
\end{abstract}


\maketitle

\section{Introduction}

We define the optical Poynting singularity as an isolated point,
line or surface at which the Poynting vector of electromagnetic beam
vanishes \cite{NovPRA}. This singularity is understood in the sense
of indeterminate direction of the Poynting vector \cite{Bekshaev07}.
In Refs. \cite{Mokhun1, Mokhun2, Mokhun3, Mokhun4} I. Mokhun et al
use another definition of the Poynting vector singularity, according
to which transversal component vanishes. This definition can be
preferable, because Poynting vector does not achieve exact zero at
any point of the field \cite{Mokhun3}. It is evident that Mokhun's
definition of the singularity includes our definition. In other
words, we are to study only for a part of singular points which
appear in terms of transverse component singularity.

The most well-known type of singularity is the phase one. It arises,
when the points of zero Poynting vector are at the same time the
points, where electric or magnetic field (or both) is zero, so that
the field phase cannot be defined \cite{Nye}. It seems to be the
typical singularity for scalar beams and finds a use for
super-resolution \cite{D'Aguanno, Perez, Tychin}, or construction
and propagation of optical vortices \cite{D'Aguanno, Indeb}. Another
well-known type of singularity is the polarization singularity (see
theoretical \cite{NyeHaj, Freund01, Freund07,Angelsky1} and
experimental \cite{Angelsky2} results), when the state of
polarization (in general, elliptical) is not fully determined. These
singularities include the points of linear polarization, or L-points
(rotation direction of polarization is not defined), and the points
of circular polarization, or C-points (polarization azimuth is not
defined). At L- and C-points the Poynting vector is not zero in
general. In some special situations, the energy flux density may
vanish owing to the certain directions of the vectors of electric
and magnetic fields. It can happen at the points of linear or
circular polarization. However, it is not the case in general,
because vanishing of the Poynting vector due to polarization
(polarization induced Poynting singularity) can be at the points of
elliptical polarization, too. Finally, it should be noted that the
fundamental connection between polarization and Poynting
singularities was noted in Refs. \cite{Mokhun3, Mokhun4}. Both
singularities do not necessarily coincide, but go together. Such a
relation can be applied for estimation of the magnitude of the
angular momentum in some spatial region using field polarization
characteristics.

\begin{figure*}[t!]
\centering \includegraphics[scale=0.75, clip=]{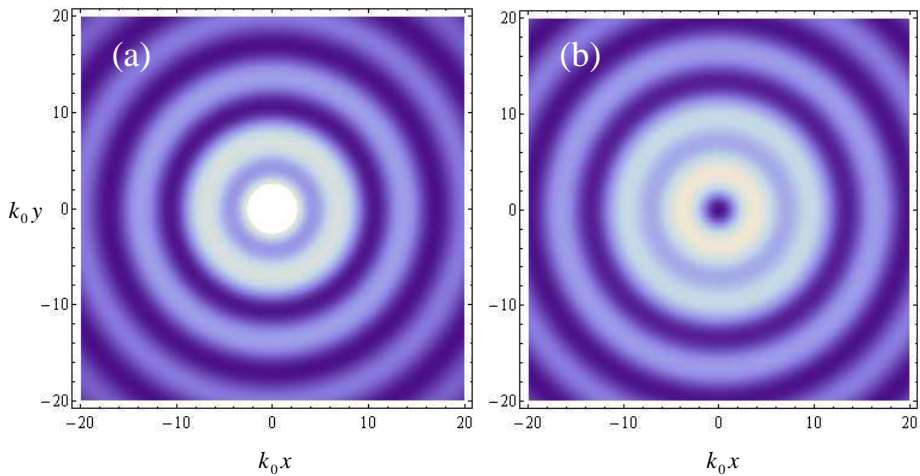}
\caption{\label{fig0} Intensity distribution (longitudinal Poynting
vector component in Eq. (\ref{flux})) of the TE-polarized vector
Bessel beam with (a) $m=1$ and (b) $m=2$. Parameters: $q/k = 0.5$
($k$ is the wavenumber in vacuum), $\varepsilon=1$, and $\mu=1$
(dielectric permittivity and magnetic permeability of the medium). }
\end{figure*}

In Ref. \cite{NovPRA}, the connection between Poynting singularities
and singular points of the dynamic systems has been developed. The
main definitions and theoretical methods related to arbitrary
electromagnetic fields have been introduced therein. In the present
paper we study the singularities of vector Bessel beams. Bessel
beams \cite{Durnin} have been studied for several decades due to the
properties of non-diffraction, self-reconstruction, and angular
momentum transfer (see, e.g., \cite{Bouchal95, Bouchal, Arlt} and
reviews \cite{Allen, McGloin} and references therein). In the
current paper we consider the exact solutions of the Maxwell
equations called vector Bessel beams. The fields of vector Bessel
beams are characterized by some polarization distribution, while the
intensity of the beam has the form depicted in Fig. \ref{fig0}. A
conventional Bessel beam follows from the vectorial one in the
paraxial approximation. Some peculiarities connected with the vector
nature of electromagnetic Bessel beams were discussed in the papers
\cite{NovJOSA2007, NovOC1, NovOC2, NovJOA}. The number of
applications which use both propagating and evanescent Bessel beams
(such as super-resolution lenses mentioned above) grows from year to
year. That is why it seems to be very important to study the
regularities of spatial structure of complex electromagnetic fields
in details. Singular peculiarities of electromagnetic beams
(especially vortices) can be used as well, for example, in optical
tweezers technique \cite{Ashkin}, detection of astronomical objects
\cite{Swartzlander}, quantum cryptography \cite{Sasada}, contrast
enhancement in microscopy \cite{Furhapter}, etc.


The special discussion is necessary on how to create a vector Bessel
beam. The field with such intensity and polarization distributions
arises in the core of the ordinary circular fibers. If the core
radius is great, the intensity distribution contains a lot of minima
and maxima. The field getting out the fiber can be considered as the
vector Bessel beam. The transverse wavenumber $q$ of such a beam is
defined by the dispersion equation.

The paper consists of introduction, two sections and conclusion.
Section 2 is devoted to propagating vector Bessel beams: we give the
basic information about their origin from Maxwell's equations and
consider the conditions of singularity generation. Section~3 deals
with the evanescent beams which may arise under the conditions of
total internal reflection and attenuate when propagating. Both
sections 2 and 3 include the discussion of electromagnetic field
properties at singular points (in particular, the state of
polarization) and the identification of the singularity type.

\section{Singularities of propagating Bessel beams}

The $m$-th order Bessel beam solution of Maxwell's equations can be
written in cylindrical coordinates ($r$, $\varphi$, $z$) as
\begin{equation}
\left( \begin{array}{c} \ee (\rr, t) \\ \hh (\rr, t) \end{array}
\right) = \left( \begin{array}{c} \ee (r, \phi) \\ \hh (r, \phi)
\end{array} \right) \exp(\rmi \beta z + \rmi m \phi - \rmi \omega
t),
\end{equation}
where longitudinal field components (along beam's axis) satisfy the
Bessel equation
\begin{equation}
\frac{{\rm d}^2}{{\rm d} r^2} \left( \begin{array}{c} E_z \\ H_z
\end{array} \right) + \frac{1}{r} \frac{{\rm d}}{{\rm d} r}
\left( \begin{array}{c} E_z \\ H_z \end{array} \right) + (q^2 -
\frac{m^2}{r^2}) \left( \begin{array}{c} E_z \\ H_z
\end{array} \right) = 0. \label{bessel_eq}
\end{equation}
Here $\omega$ is the circular frequency of electromagnetic wave,
$\beta$ is the longitudinal wavenumber,
$q=\sqrt{k^2\varepsilon\mu-\beta^2}$ is the transverse (radial)
wavenumber, $m$ is an integer number. Assuming finite-valued
solution of equation (\ref{bessel_eq}), it can be written as
follows:
\begin{equation}
H_z= c_1 J_m (q r), \qquad E_z= c_2 J_m (q r), \label{longit_comp}
\end{equation}
where $c_1$ and $c_2$ are arbitrary complex numbers. Using these
expressions for the longitudinal components, one can derive the
other components of the fields \cite{NovJOSA2007,NovOC1}:
\begin{eqnarray}
E_r&=& -\frac{m \mu k}{q^2 r} c_1 J_m (q r) + \rmi\frac{\beta}{2 q}
c_2 (J_{m-1}(q r) - J_{m+1}(q r)), \nonumber \\
E_\phi&=&-\rmi\frac{\mu k}{2 q} c_1 (J_{m-1}(q r) - J_{m+1}(q r)) -
\frac{m \beta}{q^2 r} c_2 J_{m}(q r), \nonumber \\
H_r&=&\rmi\frac{\beta}{2 q} c_1 (J_{m-1}(q r) - J_{m+1}(q r)) +
\frac{m \varepsilon k}{q^2 r} c_2 J_m (q r), \nonumber \\
H_\phi&=& - \frac{m \beta}{q^2 r} c_1 J_{m}(q r) +
\rmi\frac{\varepsilon k}{2 q} c_2 (J_{m-1}(q r) - J_{m+1}(q r)).
\label{transverse_comp}
\end{eqnarray}
where $k = \omega / c$ is the wavenumber in vacuum, $c$ is the speed
of light. The meaning of the parameters $c_1$ and $c_2$ becomes
clear, if we turn one of them to zero. If $c_1=0$, then $H_z=0$ and
the general expressions (\ref{longit_comp}) and
(\ref{transverse_comp}) are reduced to the fields of TM-polarized
beam. Otherwise ($c_2=0$) TE-polarized Bessel beam is obtained. It
is very important that the amplitudes $c_1$ and $c_2$ are complex
numbers which describe all possible vector Bessel beams of the order
$m$. In contrast to commonly used scalar beams, one of the most
important properties of these vector (electromagnetic) beams is
polarization.

The information to be discussed below is connected with the energy
flux density (Poynting vector) of the beam considered. By
substituting field components (\ref{longit_comp}) and
(\ref{transverse_comp}) into ${\s}=(c/8 \pi) {\rm Re}(\ee \times
\hh^\ast)$, the quantity ${\s}$ can be derived as
\begin{widetext}
\begin{eqnarray}
\s &=& \frac{c}{8 \pi} \left( \frac{k \beta}{q^2} (\mu |c_1|^2 +
\varepsilon |c_2|^2) (J_m^{\,\prime\, 2} + \frac{m^2}{q^2 r^2}
J_m^2) - \frac{2 m}{q^3 r} (\beta^2 + k^2 \varepsilon \mu) {\rm
Im} (c_1 c_2^\ast) J'_m J_m \right) \ez \nonumber \\
&+& \frac{c}{8 \pi} \left( \frac{m k}{q^2 r} (\mu |c_1|^2 +
\varepsilon |c_2|^2) J_m^2 - \frac{2 \beta}{q} {\rm Im}(c_1
c_2^\ast) J'_m J_m \right) \ep, \label{poynting_p}
\end{eqnarray}
\end{widetext}
where the derivative is calculated with respect to the entire
function argument, i.e. $J'_m (qr)={\rm d} J_m/{\rm d}(q r)$. This
expression describes the dependence of Poynting vector on radial
coordinate $r$ for vector Bessel beam with any particular value of
$m$ which is a discrete parameter. This means that if the condition
$m=0$ is taken (zeroth order beam), all terms with $m$ in numerator
vanish and there is no any ambiguities (and discontinuities) at the
point $r=0$.

Two situations, $m=0$ and $m \neq 0$, are different from the
mathematical point of view. Therefore, we will consider them
separately. In the case $m=0$, the expression for Poynting vector
(\ref{poynting_p}) is reduced to
\begin{eqnarray}
S_r(r)&=&0, \nonumber \\
S_\phi(r)&=&-\frac{c \beta}{4\pi q} {\rm Im}(c_1
c_2^*) J_{0}(qr) J_{1}(qr), \label{flux_m0} \\
S_z(r)&=&\frac{c k\beta}{8 \pi q^2} (\mu |c_1|^2+\varepsilon
|c_2|^2) J_{1}^2(qr). \nonumber
\end{eqnarray}
It is seen that, though the beam as a whole is directed along
$z$-axis, the Poynting vector is not. This is due to vector nature
of the beam considered which is a general solution of Maxwell's
equations. Taking only TE- or TM-component ($c_1$ or $c_2$ is null)
and proceeding to paraxial limit, one can obtain the usual (scalar)
result with the only $S_z$ component. Note, that in Eqs.
(\ref{flux_m0}) the first expression stands for the property of
diffractionless of the Bessel beam, while the second one corresponds
to angular momentum transfer by the beam. Finally, it is worth to
stress that the straight physical sense of energy has not the
Poynting vector itself, but the integral of it over the whole cross
section of the beam. This integral is always directed along
propagation of the beam as a whole (say, $z$-axis). But the Poynting
vector of a complex beam (which consists of enormous number of plane
waves) at a certain point can be directed in various ways (including
opposite direction, $-z$) \cite{Kats, NovJOSA2007}.

The position of an optical Poynting singularity $\rr_0$ can be found
from equation ${\s}(\rr_0)=0$ or, in the case of propagating vector
Bessel beam,
\begin{eqnarray}
S_\phi(r_0)=0, \qquad S_z(r_0)=0. \label{system}
\end{eqnarray}
The couple of equations (\ref{system}) is satisfied, when
\begin{equation}
f(r_0) = J_1(q r_0)=0. \label{f_m=0}
\end{equation}
The positions of singularities $r_0$ are defined by the single beam
parameter $q$. For paraxial beams (transverse wavenumber $q$ is
small compared with the wavenumber), the radial coordinates $r_0$
are much greater than that for non-paraxial vector beams. It is
important that the singularities exist for any $c_1$ and $c_2$. This
feature is emphasized in contrast to what will be discussed below
for $m \neq 0$.

If $m \neq 0$, then using the formulae
\begin{eqnarray}
J'_m (qr)&=&\frac{1}{2} \left( J_{m-1}(qr) - J_{m+1}(qr) \right),
\nonumber \\ J_m &=& \frac{q r}{2 m} \left( J_{m-1} + J_{m+1}
\right), \label{BFproperties}
\end{eqnarray}
the Poynting vector components (\ref{poynting_p}) can be represented
in the form
\begin{widetext}
\begin{eqnarray}
S_r(r)&=&0, \nonumber \\
S_\phi(r)&=&\frac{c r}{32 m \pi} \left[ k(\mu |c_1|^2+\varepsilon
|c_2|^2)(J_{m-1}+J_{m+1})^2-2\beta {\rm Im}(c_1
c_2^*)(J_{m-1}^2-J_{m+1}^2) \right],  \label{flux}\\
S_z(r)&=&\frac{c}{8 \pi} \left[ \frac{k\beta}{2q^2}(\mu
|c_1|^2+\varepsilon
|c_2|^2)(J_{m-1}^2+J_{m+1}^2)-\frac{\beta^2+k^2\varepsilon\mu}{2q^2}
{\rm Im}(c_1 c_2^*)(J_{m-1}^2-J_{m+1}^2) \right]. \nonumber
\end{eqnarray}

It leads to the two systems of equations (with upper and lower
signs):
\begin{eqnarray}
&f&(r_0) \equiv (-\beta \pm
k\sqrt{\varepsilon\mu})J_{m-1}(r_0)+(\beta \pm
k\sqrt{\varepsilon\mu})J_{m+1}(r_0)=0, \label{eq1} \\
&\mu& |c_1|^2+\varepsilon |c_2|^2 \mp 2 \sqrt{\varepsilon\mu}
\rm{Im}(c_1 c_2^*)=0. \label{eq2}
\end{eqnarray}
\end{widetext}

The solutions of equation (\ref{eq1}) are $r_0$ at different
$\beta\geq0$ and $m$. Positions of singular points are determined by
the only coordinate $r_0$ in three-dimensional space. Therefore, the
singularities are located on the cylindrical surfaces of certain
radius. Reduced to one dimension, the surface can be considered as
the point in the radial direction. The quantity of these surfaces is
infinite due to infinite number of solutions of equation
(\ref{eq1}). The choice of sign in equation (\ref{eq1}) is connected
with the choice of sign in relation (\ref{eq2}) which defines the
coefficients $c_1$ and $c_2$ and, hence, the Bessel beam itself.
Equation (\ref{eq2}) can be represented as quadratic form due to the
common representation of complex numbers $c_1=a_1
\rm{e}^{\rm{i}\varphi_1}$ and $c_2=a_2 \rm{e}^{\rm{i}\varphi_2}$
($a_1$ and $a_2$ are real and positive):
\begin{eqnarray}
\mu a_1^2+\varepsilon a_2^2 \mp 2 \sqrt{\varepsilon\mu} a_1 a_2
\sin{\Delta\varphi}=0, \label{quadreq}
\end{eqnarray}
where $\Delta\varphi=\varphi_1-\varphi_2$. Expressing the phase
difference $\Delta\varphi$, we have
\begin{eqnarray}
\sin{\Delta\varphi}=\pm \frac{\mu a_1^2+\varepsilon a_2^2}{2
\sqrt{\varepsilon\mu} a_1 a_2}. \label{sinphi}
\end{eqnarray}
Taking into account that for $a_1$, $a_2\geq0$ the inequality $\mu
a_1^2+\varepsilon a_2^2\geq2 \sqrt{\varepsilon\mu} a_1 a_2$ holds,
it should be stated that $|\sin{\Delta\varphi}|\geq1$. Owing to the
limitation on value of sine equation (\ref{sinphi}) can be reduced
to $\sin{\Delta\varphi}=\pm1$, so that
\begin{eqnarray}
a_1=\sqrt{\frac{\varepsilon}{\mu}} a_2, \qquad \Delta\varphi=\pm
\frac{\pi}{2}+2 \pi p, \label{a1}
\end{eqnarray}
where $p$ is integer. Thus, we conclude that vector Bessel beam can
contain the singularities in beam's cross-section, only if it is
constructed of TE- and TM-components oscillating out of phase and
possessing amplitudes matched by the wave impedance $\sqrt{\mu /
\varepsilon}$. The simplest case of the relation (\ref{a1}) (for
propagation in vacuum) gives $c_1=1$, $c_2=-\rmi$ and $c_1=1$,
$c_2=\rmi$ for upper-sign and lower-sign equations (\ref{a1}),
respectively.

\begin{figure}[t!]
\centering \includegraphics[scale=0.85, clip=]{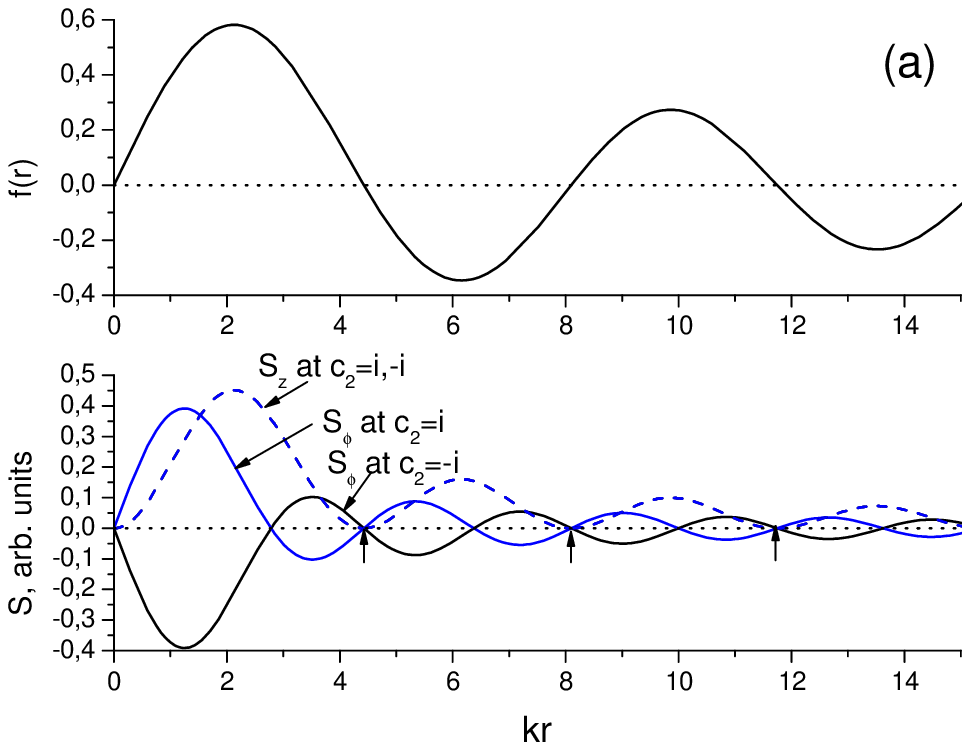}
\includegraphics[scale=0.85, clip=]{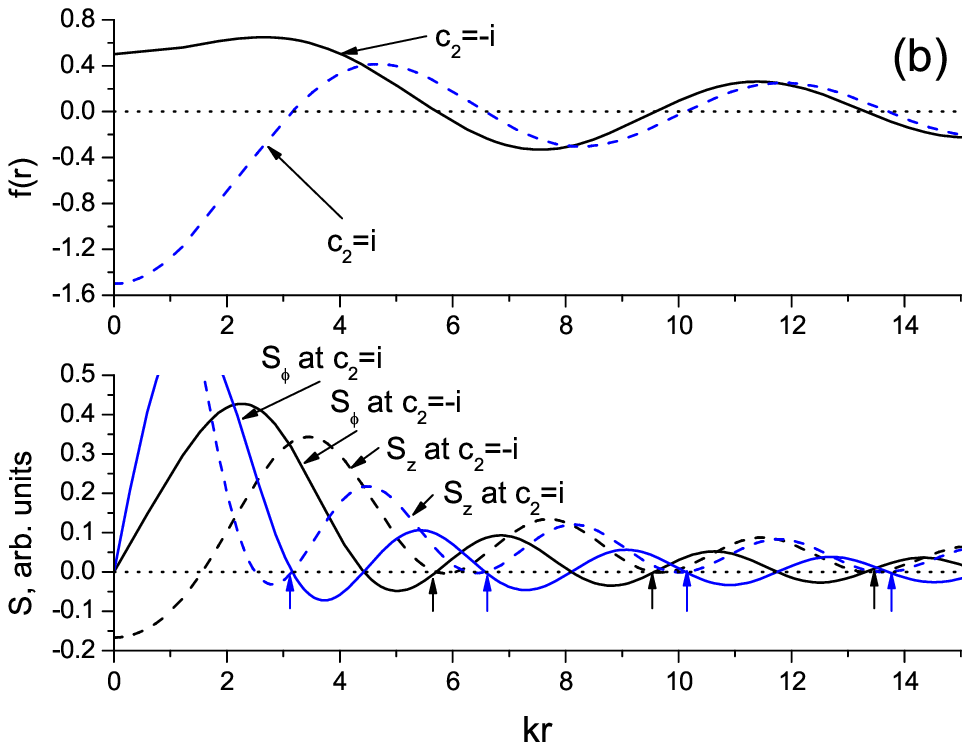}
\caption{\label{fig1} Function $f(r)$ (in the upper figures) and the
components of Poynting vector (in the bottom figures) for (a) $m=0$
and (b) $m=1$. Function $f$ is defined as the left-hand side of
equation (\ref{f_m=0}) for $m=0$ or equation (\ref{eq1}) for $m=1$.
Selecting parameter $c_1=1$ for $m=1$, the another parameter
$c_2=\mp \rmi$ follows from the choice of sign in equation
(\ref{eq1}). Vertical arrows in the bottom figures mark the radii of
singularities. Other parameters: $\beta=0.5k$, $\varepsilon=1$,
$\mu=1$.}
\end{figure}

As examples, we consider vector Bessel beams of the orders $m=0$ and
$m=1$. In figure \ref{fig1} we graphically solve equation
(\ref{f_m=0}) for $m=0$ and equation (\ref{eq1}) for $m=1$. The zero
crossings of the function $f(r)$ (left-hand sides of equations
(\ref{f_m=0}) and (\ref{eq1})) in the upper figures correspond to
the vertical arrows in the bottom figures. In figure \ref{fig1}(b)
the most of Poynting singularities for the amplitudes $c_2=\rmi$ and
$c_2=-\rmi$ appear together. The single exception is the first
singular point for $c_2=\rmi$. Spatial distance between a couple of
Poynting singularities diminishes with increasing the radial
coordinate $r$, because the components of the energy flux densities
are similar for great $r$. In figure \ref{fig1}(a), $f(r)$-curves
for $c_2=\rmi$ and $c_2=-\rmi$ (as well as for any amplitudes $c_1$
and $c_2$) coincide.


\begin{figure}[t!]
\centering \includegraphics[scale=0.85, clip=]{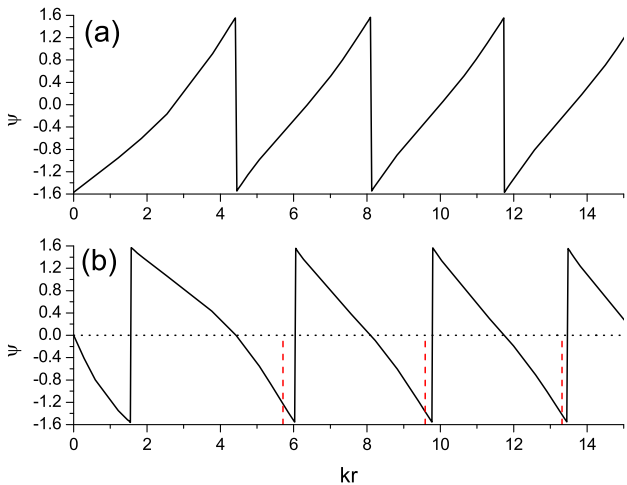}
\caption{\label{fig2} Angle $\psi$ vs. radial coordinate $r$ for (a)
$m=0$ and (b) $m=1$. Vertical dashed lines mark the radii of
singularities. Parameters: $\beta=0.5k$, $\varepsilon=1$, $\mu=1$,
$c_1=1$, $c_2=-\rmi$.}
\end{figure}


Since the radial component of the Poynting vector $S_r$ is absent
(see equation (\ref{flux})), the direction of vector $\s$ in the
plane ($\varphi$, $z$) can be specified by the angle
$\psi(r)=\arctan[S_\phi(r)/S_z(r)]$.

Figure \ref{fig2}(a) shows the change of the angle $\psi(r)$ with
radial coordinate. Its value oscillates between $-\pi$ and $\pi$,
the function $\psi$ being discontinuous. The function jumps from
$\pi$ to $-\pi$. For $m=0$, the positions of jumps $r_j$ coincide
with that of singularities. In fact, at the singularities the
expression for $\psi(r)$ contains an ambiguity of the $0/0$ type,
which can be unwound as
\begin{eqnarray}
\psi(r_0)=\arctan S'_\phi(r_0)/S'_z(r_0) \label{psi_r0},
\end{eqnarray}
where $S'_\phi(r_0)$ and $S'_z(r_0)$ are the derivatives of the
Poynting vector components at the singular point $r_0$. For $m=0$ we
get
\begin{eqnarray}
S'_\phi (r_0)=- \frac{c}{4 \pi} \beta {\rm Im} (c_1 c_2^*) J_0^2 (q
r_0), \qquad S'_z (r_0)=0, \nonumber
\end{eqnarray}
so that $\psi(r_0)$ equals $\pi$, i.e. singularity points coincide
with points of jumps $r_j$. At the singularity only one of
components of Poynting vector (namely, $S_\phi$) changes the sign.
Such a situation can occur, if $S_z$ is tangent to the abscissa axis
at singularity. This statement is confirmed by the figure
\ref{fig1}(a) for the Poynting vector components.

The function $\psi(r)$ for $m=1$ is shown in figure \ref{fig2}(b).
For some $r$, the jumps from $-\pi$ to $\pi$ occur. The positions of
jumps $r_j$ are determined by vanishing $S_z$ (at the same time
$S_{\phi} \neq 0$). The radial coordinates of singularities $r_0$
(see vertical dashed lines in figure \ref{fig2}(b)) do not coincide
with $r_j$. It should be noted that at the points $r_0$ both
components of Poynting vector, $S_z$ and $S_\phi$, simultaneously
reverse sign. In figure \ref{fig2}(b) we observe that $\psi(r_0)<0$,
therefore, if $S_\phi>0$ and $S_z<0$ on the left of the singularity,
then $S_\phi<0$ and $S_z>0$ on the right of it. The distance between
$r_0$ and $r_j$ gets smaller as radius increases, whereas the angle
$\psi$ converges to $-\pi$ (or $\pi$ for another value of $m$).

It should be noted that considered solutions for the vector Bessel
beams are the centrally symmetric beams. They are very attractive
due to the existence of the closed-form expressions for the fields.
On the other hand, the studied beams are very specific, because of
different imperfections during their realistic generation. So, if we
introduce a small electric field ${\bf e}(\rr)$ as a fluctuation,
the field can be written as ${\bf E} = {\bf E}_B + {\bf e}$ and
${\bf H} = {\bf H}_B + {\bf h}$, where ${\bf E}_B$ and ${\bf H}_B$
are the electric and magnetic fields of the ideal vector Bessel
beam, and ${\bf h} = 1/(\rmi k \mu) \nabla \times {\bf e}$. The
fluctuation deforms the beam form, so that it becomes asymmetric.
Then the Poynting vector takes the form
\begin{equation}
{\bf S}(\rr) \approx {\bf S}_B(r) + {\bf s}(\rr), \qquad {\bf s} =
\frac{c}{4 \pi} {\rm Re}\left( {\bf E}_B \times {\bf h} + {\bf e}
\times {\bf H}_B \right).
\end{equation}
It is evident that after adding ${\bf s}$ the singular point $r_0$
will be destroyed. Since the fluctuation field is an arbitrary one,
it is quite impossible to meet the condition ${\bf S}(\rr)=0$ now
(only for some specific fields ${\bf e}$ it can be done). However,
as the field ${\bf e}(\rr)$ is small, the singularities appear to be
approximately situated at $r_0$, however ${\bf S}(\rr)=0$ is not
satisfied exactly at any point.

Further we consider the electric and magnetic fields at the singular
surfaces. None of them vanish as it would be in the case of phase
singularity. In the case $m=0$ the azimuthal components vanish and
both electric and magnetic fields are directed along $\ez$, while
$c_1$ and $c_2$ are arbitrary complex numbers, i.e.
\begin{eqnarray}
\ee(r, \phi)=c_2 J_0 (q r_0) \ez, \qquad \hh(r, \phi)=c_1 J_0 (q
r_0) \ez.
\end{eqnarray}
Such longitudinal fields are linearly polarized and can be treated
as polarization L-singularities.

For the rest of $m$s, one can derive the fields at singular points
from equations (\ref{longit_comp}) and (\ref{transverse_comp}):
\begin{eqnarray}
\ee(r_0, \phi)&=&-c_2 \frac{\beta \pm k \sqrt{\varepsilon \mu}}{q}
J_{m+1}(q r_0) \ep \nonumber \\ &+& c_2 J_m (q r_0) \ez, \nonumber \\
\hh(r_0, \phi)&=&-c_1 \frac{\beta \pm k \sqrt{\varepsilon \mu}}{q}
J_{m+1} (q r_0) \ep \nonumber \\ &+& c_1 J_m (q r_0) \ez.
\label{fieldsin}
\end{eqnarray}
We may note that the magnetic and electric field strengths are
connected as
\begin{eqnarray}
\hh=\frac{c_1}{c_2} \ee. \label{field_connect}
\end{eqnarray}
By substituting the coefficients $c_1$ and $c_2$ from equation
(\ref{a1}), we represent the relation (\ref{field_connect}) in the
form $\hh=\pm \rmi \sqrt{\varepsilon/\mu} \ee$.


\begin{figure}[t!]
\centering \includegraphics[scale=0.85, clip=]{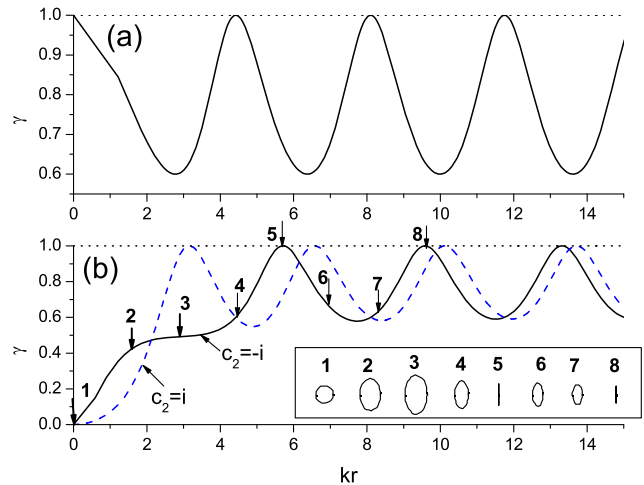}
\caption{\label{fig3} Polarization parameter $\gamma$ versus radial
coordinate for (a) $m=0$, $c_2=-\rmi$, (b) $m=1$. The inset depicts
the polarization ellipses at the points of solid curve marked by
vertical arrows. Parameters: $c_1=1$, $\beta=0.5k$, $\varepsilon=1$,
$\mu=1$.}
\end{figure}


Thus, the singularities considered are due to the special link
between the electric and magnetic fields. We call them the
polarization induced Poynting singularities. In general, they are
distinct from the polarization singularities, which arise at the
points of linear (L-points) or circular (C-points) polarizations
\cite{Freund01,Freund07}.

The state of polarization can be studied by means of parameter
$\gamma$ \cite{Fedorov}:
\begin{eqnarray}
\gamma=\frac{|\ee^2|}{|\ee|^2}. \label{gamma}
\end{eqnarray}
The introduced parameter characterizes the polarization of an
electromagnetic beam using the complex electric field vector. The
same analysis can be performed for the polarization of the magnetic
field. All the types of polarization can be described with $\gamma$.
So, circular polarization corresponds to $\gamma=0$, elliptical one
--- to $0<\gamma<1$, linear polarization --- to $\gamma=1$. In
figure \ref{fig3} the singularities are located at the maxima of
curves $\gamma(r)$. That is why the electromagnetic fields at the
singular surfaces are linearly polarized ($\gamma=1$). The prove of
$\gamma=1$ can be fulfilled analytically: the value $\gamma$ for the
fields (\ref{fieldsin}) equals $|c_2^2|/|c_2|^2=1$. The inset of the
figure \ref{fig3}(b) clearly demonstrates the evolution of the forms
of polarization ellipses. Starting with circular polarization at
beam's axis ($r=0$), it changes step-by-step through elliptical to
linear one at singularity $r_0$. It should be noted that circular
polarization appears only at beam's axis. Hence the axis is the
place of C-points. Moreover, at $r=0$ the fields can be equal to
zero (e.g., for $m=2$), i.e. the phase singularity exists at this
point. So, the beam axis simultaneously corresponds to phase,
polarization, and Poynting singularities. In this situation, the
Poynting singularity is caused by the field vanishing, but not by
the special polarization. However, the other Poynting singularities
of the vector Bessel beam are the polarization induced ones.
Sometimes, they can be identified as L-points according to the
classification of polarization singularities. In our example, the
Poynting singularities correspond to the L-points. In general, such
is not the case, because the definitions of both singularities are
not connected directly. For $m=0$ the beam axis $r_0=0$ is also the
singular line (see figure \ref{fig3}a), the electric and magnetic
fields being non-zero ones and the Poynting singularity being caused
by the linear field polarization. The link between polarization and
Poynting singularities for arbitrary electromagnetic fields is
discussed in Refs. \cite{Mokhun3, Mokhun4}.

\section{Poynting singularities for evanescent beams}

When a vector Bessel beam falls onto the interface between two
media, the total internal reflection can occur. The Bessel beam that
penetrates the interface exponentially decays, when moving from the
boundary. Such a wave is called the evanescent (inhomogeneous) one.
Thus, the evanescent Bessel beam corresponds to the situation when
$q>k \sqrt{\varepsilon \mu}$. Applying replacement $\beta
\rightarrow \rmi \beta'=\rmi \sqrt{q^2-k^2 \varepsilon \mu}$ to the
fields (\ref{longit_comp}) and (\ref{transverse_comp}) we find the
Poynting vector components \cite{NovJOA}:
\begin{eqnarray}
S_r&=&-\frac{c \beta' m}{4\pi q^2 r} {\rm
Im}(c_1 c_2^*) J_m^2(q r) {\rm e}^{-2 \beta' z}, \nonumber \\
S_\phi&=&\frac{c k m}{8\pi q^2 r} (\mu |c_1|^2+\varepsilon
|c_2|^2)J_m^2(q r) {\rm e}^{-2 \beta' z}, \label{flux_evan} \\
S_z&=&- \frac{c m (\beta'^2+k^2\varepsilon\mu)}{4 \pi q^3 r} {\rm
Im}(c_1 c_2^*)J_m(q r) J'_m(q r) {\rm e}^{-2 \beta' z}. \nonumber
\end{eqnarray}
In contrast to the propagating beam (see equation
(\ref{poynting_p})), the evanescent Bessel beam has the radial
component $S_r$, i.e. it is diffracted. As in previous section, one
can mark out two cases, $m = 0$ and $m \neq 0$.

For $m=0$, it is turned out that the Poynting vector $\s$ vanishes
for each radial coordinate $r$. However, neither electric nor
magnetic field is equal to zero. The fields are expressed as
\begin{widetext}
\begin{eqnarray}
\ee(\rr) = {\rm e}^{-\beta' z} \left( \frac{\beta'}{q} c_2 J_1(q r)
\er + \rmi \frac{\mu k}{q} c_1 J_1(q r) \ep + c_2 J_0(q r) \ez \right),  \nonumber\\
\hh(\rr) = {\rm e}^{-\beta' z} \left( \frac{\beta'}{q} c_1 J_1(q r)
\er - \rmi \frac{\varepsilon k}{q} c_2 J_1(q r) \ep  + c_1 J_0(q r)
\ez \right). \label{fields_evan_m0}
\end{eqnarray}
\end{widetext}
We have obtained the very exciting situation. There exist non-zero
cylindrically symmetric solutions of the Maxwell equations, for
which the energy is not transferred (the energy flux is entirely
zero). Such a situation is similar to the case of the plane-wave
standing waves. Being the Beltrami-fields, these standing waves do
not transfer the energy, too \cite{Zaghloul}.

Now the singularities are not located at certain cylindrical
surfaces. They occupy the whole three-dimensional space. So, the
singularity can be not only the isolated point, or line, or surface,
but the space itself. In general, the polarization of
electromagnetic field (\ref{fields_evan_m0}) is elliptical. Exactly
such a field polarization provides the complete vanishing of the
energy flux. At the same time, polarization is linearly or
circularly polarized at some points. For example, L-points arise,
when the condition $J_1 (q r) = 0$ holds true. It is obvious, that
Poynting singularities include the polarization singularities, which
are situated at the cylindrical surfaces.

\begin{figure}[t!]
\centering \includegraphics[scale=0.9, clip=]{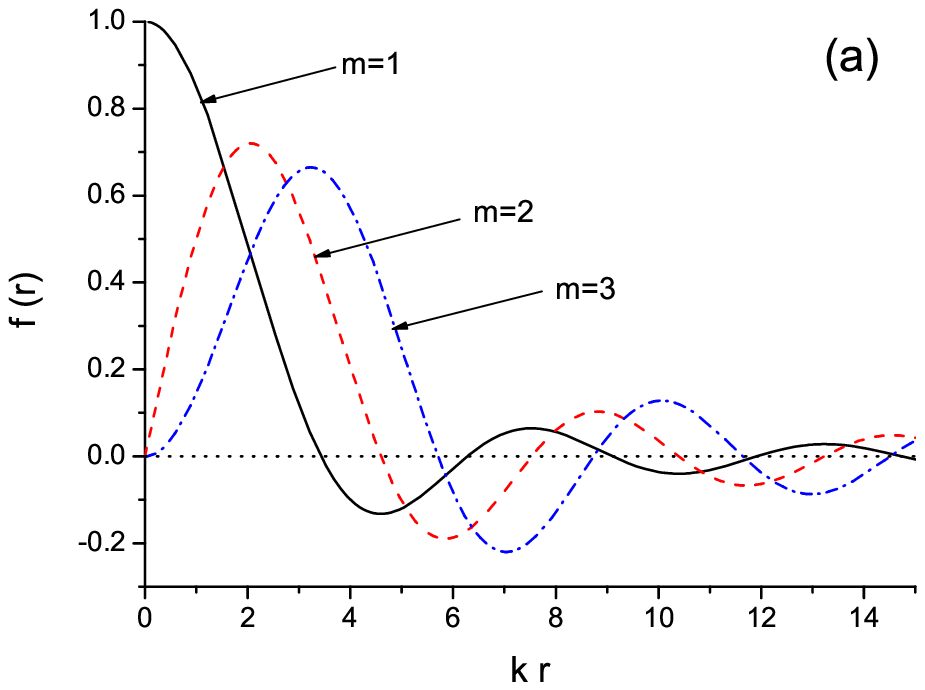}
\includegraphics[scale=0.85, clip=]{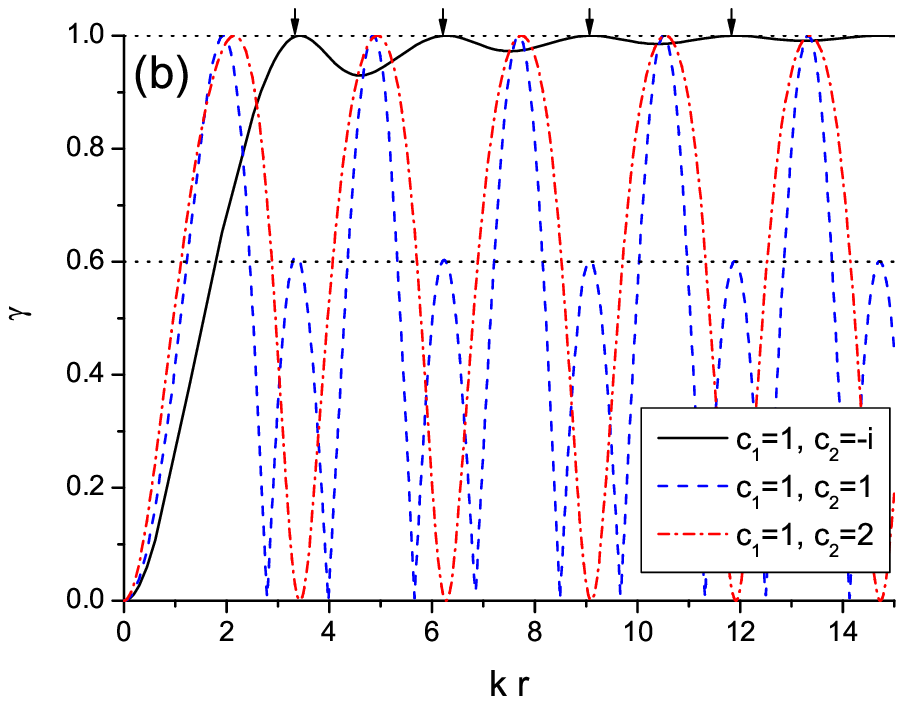}
\caption{\label{fig4} (a) Function $f(r)$ (left-hand side of
equation (\ref{eq_evan})) for different values of beam order $m$.
(b) Polarization parameter $\gamma$ versus radial coordinate for
different amplitudes $c_1$ and $c_2$, and $m=1$. Arrows stand for
singularity radii $r_0$. Parameters: $\beta'=0.5k$, $\varepsilon=1$,
$\mu=1$.}
\end{figure}

For $m \neq 0$, the Poynting vector components (\ref{flux_evan}) can
be rewritten using the properties of Bessel functions
(\ref{BFproperties}):
\begin{eqnarray}
S_r&=& -\frac{c \beta' r}{16 \pi m} {\rm
Im}(c_1 c_2^*) (J_{m-1}+J_{m+1})^2 {\rm e}^{-2 \beta' z}, \label{flux_evan_m} \\
S_\phi &=& \frac{c k r}{32 \pi m} (\mu |c_1|^2+\varepsilon
|c_2|^2)(J_{m-1}+J_{m+1})^2 {\rm e}^{-2 \beta' z}, \nonumber \\
S_z &=& -\frac{c(\beta'^2+k^2\varepsilon\mu)}{16 \pi q^2} {\rm
Im}(c_1 c_2^*)(J_{m-1}^2-J_{m+1}^2) {\rm e}^{-2 \beta' z}. \nonumber
\end{eqnarray}
The singularity position $r_0$ is determined from the equation
\begin{eqnarray}
f(r_0) \equiv J_{m-1}(q r_0) + J_{m+1}(q r_0) = 0. \label{eq_evan}
\end{eqnarray}
The above equation holds for any complex coefficients $c_1$ and
$c_2$. This means that any evanescent Bessel beam has singular
points or, rather, cylindrical surfaces. In figure \ref{fig4}(a) the
left-hand side of equation (\ref{eq_evan}) is shown. For $m>1$ the
Poynting singularity at beam's axis arises. It is caused by zero
fields and coincides with the phase singularity as for the
propagating electromagnetic Bessel beams.


As to the electric and magnetic fields at singular points, they take
the form
\begin{eqnarray}
\ee (\rr) = {\rm e}^{\rmi m \varphi - \beta' z} J_{m+1}(q r_0)
\left( \frac{\beta'}{q} c_2 \er + \rmi \frac{\mu k}{q} c_1 \ep
\right), \nonumber \\
\hh (\rr) = {\rm e}^{\rmi m \varphi - \beta' z} J_{m+1}(q r_0)
\left( \frac{\beta'}{q} c_1 \er - \rmi \frac{\varepsilon k}{q} c_2
\ep \right). \label{filds_evan}
\end{eqnarray}
It is important that the fields satisfy the relation
\begin{eqnarray}
\hh=\alpha_1 \ee + \alpha_2 \ee^\ast, \label{fieldcon}
\end{eqnarray}
where $\alpha_1$ and $\alpha_2$ are real and complex numbers,
respectively. The numbers are equal to
\begin{eqnarray}
\alpha_1=\frac{\mu |c_1|^2 - \varepsilon |c_2|^2}{2\mu {\rm Re}(c_1
c_2^\ast)}, \qquad \alpha_2=\frac{\mu c_1^2 + \varepsilon
c_2^2}{2\mu {\rm Re}(c_1 c_2^\ast)}. \nonumber
\end{eqnarray}
Relation between the electric and magnetic fields (\ref{fieldcon})
is necessary to turn the Poynting vector to zero. Generally, the
field link (\ref{fieldcon}) corresponds to the elliptical
polarization. The state of polarization can be described by the
parameter $\gamma$ (see its definition (\ref{gamma})). For
evanescent Bessel beams, we will express it in the closed form as
\begin{eqnarray}
\gamma&=&\frac{|\beta'^2c_2^2-\mu^2 k^2 c_1^2|}{\mu^2 k^2
|c_1|^2+\beta'^2 |c_2|^2} \nonumber \\ &\equiv& \sqrt{1 - \frac{2
\mu^2 k^2 \beta'^2 \left[ |c_1|^2 |c_2|^2 + {\rm Re}(c_1^2 c_2^{\ast
2}) \right]}{(\mu^2 k^2 |c_1|^2 + \beta'^2 |c_2|^2)^2}}.
\label{gamma_evan}
\end{eqnarray}
Figure \ref{fig4}(b) demonstrates three possible polarizations at
the singularity point $r_0$ indicated by the vertical arrows. The
linear polarization (L-point) is achieved for $\gamma=1$, i.e. for
\begin{eqnarray}
|c_1|^2 |c_2|^2 + {\rm Re}(c_1^2 c_2^{\ast 2})=0.
\end{eqnarray}
Using the above introduced representation of the complex amplitudes
$c_1=a_1 \rm{e}^{\rm{i}\varphi_1}$ and $c_2=a_2
\rm{e}^{\rm{i}\varphi_2}$ one can easily obtain the phase difference
\begin{eqnarray}
\Delta\varphi \equiv \varphi_1-\varphi_2 = \frac{\pi}{2}+\pi p,
\qquad p = 0, \pm 1, \pm 2 \ldots . \label{evan_L}
\end{eqnarray}
In figure \ref{fig4}(b) this condition is fulfilled for the
amplitudes $c_1=1$ and $c_2=-\rmi$.

Circularly polarized fields at singularities (C-points) are
described by $\gamma=0$. From equation (\ref{gamma_evan}) it follows
the necessary condition to obtain C-point at the Poynting
singularity:
\begin{eqnarray}
\beta'^2c_2^2-\mu^2 k^2 c_1^2=0.
\end{eqnarray}
By substituting the amplitudes, we get
\begin{eqnarray}
\left(\frac{a_2}{a_1}\right)^2=\left(\frac{\mu k}{\beta'}\right)^2
{\rm e}^{2 \rmi \Delta\varphi}.
\end{eqnarray}
Since $a_1$ and $a_2$ are real, the condition $\exp{(2 \rmi
\Delta\varphi)=1}$ should be valid. Hence,
\begin{eqnarray}
\Delta\varphi=\pi p, \qquad p = 0, \pm 1, \pm 2 \ldots
\end{eqnarray}
and the relation between coefficients $c_1$ and $c_2$ should satisfy
the expression
\begin{eqnarray}
\frac{c_2}{c_1}=\pm \frac{\mu k}{\beta'}. \label{evan_C}
\end{eqnarray}
Such C-singularities are shown in figure \ref{fig4}(b) as that
corresponding to the curve with amplitudes $c_1=1$ and $c_2=2$.

If neither equation (\ref{evan_L}) nor (\ref{evan_C}) holds, the
polarization of the field at the singular point is elliptical. For
example, the elliptically polarized field with $\gamma=0.6$ appears
for $c_1=c_2=1$ (see figure \ref{fig4}(b)). We can be convinced
again that the Poynting singularity for the vector Bessel beams can
include the polarization singularities as particular cases. Of
course, the polarization singularities can be generated at the
points different from the Poynting singularities, too.

\section{Conclusion}

Summing up, vector Bessel beams (solutions of Maxwell's equations in
cylindrical coordinates) demonstrate the occurrence of non-phase
optical singularities, namely Poynting singularities, which are
situated at cylindrical surfaces in three-dimensional space. To
create the singularity of the Poynting vector, TE and TM components
of propagating Bessel beams defined by the complex amplitudes $c_1$
and $c_2$ should be out of phase and be matched with the wave
impedance. Such a strong restriction on the amplitudes arises only
for propagating waves. Any evanescent Bessel beam is appeared to be
singular. Moreover, the evanescent beam with $m=0$ is singular at
any spatial point. This case describes a sort of cylindrical
standing wave. The points of Poynting singularities always include
the phase singularities as particular case (at Bessel beam's axis).
However this is not the case for L- and C-points (polarization
singularities). The manifold of the Poynting singularities and
manifold of L- and C-points can intersect, however, it is only the
coincidence. Since vector cylindrical solutions of the Maxwell
equations describe the electromagnetic modes of circular fibers, the
theory of Poynting singularities can be applied for the fields of
guiding structures.


\begin{thebibliography}{26}

\bibitem{NovPRA} A. V. Novitsky and L. M. Barkovsky, Phys. Rev. A {\bf 79},
033821 (2009).

\bibitem{Bekshaev07} A. Ya. Bekshaev and M. S. Soskin, Opt. Commun. {\bf 271}, 332 (2007).

\bibitem{Mokhun1} I. Mokhun, A. Mokhun, and Ju. Viktorovskaya, SPIE Proc. {\bf 6254}, 625409 (2006).

\bibitem{Mokhun2} I. Mokhun, A. Mokhun, and Ju. Viktorovskaya, SPIE Proc. {\bf 6254}, 625408 (2006).

\bibitem{Mokhun3} I. Mokhun, A. Mokhun, and Ju. Viktorovskaya, Ukr. J. Phys. Opt. {\bf 7}, 129 (2006).

\bibitem{Mokhun4} I. Mokhun and R. Khrobatin, J. Opt. A: Pure Appl. Opt. {\bf 10}, 064015 (2008).

\bibitem{Nye} J. F. Nye and M. V. Berry, Proc. R. Soc. A {\bf 336}, 165 (1974).

\bibitem{D'Aguanno} G. D'Aguanno, N. Mattiucci, M. Bloemer, and A. Desyatnikov, Phys. Rev. A {\bf 74},
043825 (2008).

\bibitem{Perez} M. Perez-Molina, L. Carretero, P. Acebal, and S. Blaya, J. Opt. Soc. Am. A {\bf
25}, 2865 (2008).

\bibitem{Tychin} V. P. Tychinskii, Phys. Usp. {\bf 51}, 1205 (2008).

\bibitem{Indeb} G. Indebetouw, J. Mod. Opt. {\bf 40}, 73 (1993).

\bibitem{NyeHaj} J. F. Nye and J. V. Hajnal, Proc. R. Soc. A {\bf 409}, 21 (1987).

\bibitem{Freund01} I. Freund, Opt. Commun. {\bf 199}, 47 (2001).

\bibitem{Freund07} I. Freund, Opt. Commun. {\bf 272}, 293 (2007).

\bibitem{Angelsky1} O. V. Angelsky, A. I. Mokhun, I. I. Mokhun, and M. S. Soskin, Phys. Rev. E {\bf 65}, 036602 (2002).

\bibitem{Angelsky2} O. Angelsky, A. Mokhun, I. Mokhun, and M. S. Soskin, Opt. Commun. {\bf 207}, 57 (2002).

\bibitem{Durnin} J. Durnin, J. Opt. Soc. Am. A {\bf 4}, 651 (1987).

\bibitem{Bouchal95} Z. Bouchal and M. Olivik, J. Mod. Opt. {\bf 42}, 1555 (1995).

\bibitem{Bouchal} Z. Bouchal, J. Wagner, and M. Chlup, Opt. Commun. {\bf 151}, 207
(1998).

\bibitem{Arlt} J. Arlt and M. J. Padgett, Opt. Lett. {\bf 25}, 191 (2000).

\bibitem{Allen} L. Allen, S. M. Barnett, and M. J. Padgett, Optical Angular Momentum, Institute of Physics Publishing, Bristol, 2003.

\bibitem{McGloin} D. McGloin and K. Dholakia, Cont. Phys. {\bf 46}, 15 (2005).

\bibitem{NovJOSA2007} A. V. Novitsky and D. V. Novitsky,  J. Opt. Soc. Am. A. {\bf 24}, 2844 (2007).

\bibitem{NovOC1} A. V. Novitsky and D. V. Novitsky, Opt. Commun. {\bf 281}, 2727 (2008).

\bibitem{NovOC2} A. V. Novitsky, Opt. Commun. {\bf 281}, 5310 (2008).

\bibitem{NovJOA} A. V. Novitsky and L. M. Barkovsky, J. Opt. A: Pure Appl. Opt. {\bf 10},
075006 (2008).

\bibitem{Ashkin} A. Ashkin, Biophys. J. {\bf 61}, 569 (1992).

\bibitem{Swartzlander} G. A. Swartzlander, Opt. Lett. {\bf 26}, 497 (2001).

\bibitem{Sasada} H. Sasada and M. Okamoto, Phys. Rev. A {\bf 68}, 012323 (2003).

\bibitem{Furhapter} S. Furhapter, A. Jesacher, S. Bernet, and M. Ritsch-Marte, Opt. Express {\bf 13}, 689 (2005).

\bibitem{Kats} B. Z. Katsenelenbaum, J. Commun. Technol. Electron. {\bf 42}, 119 (1997).

\bibitem{Fedorov} F. I. Fedorov, Theory of Gyrothropy, Nauka i Tekhnika, Minsk, 1976.

\bibitem{Zaghloul} H. Zaghloul and H. A. Buckmuster, Am. J. Phys. {\bf 56}, 801 (1988).

\bibitem{Lakhtakia} A. Lakhtakia, Beltrami Fields in Chiral Media, World
Scientific, Singapore, 1994.

\end{thebibliography}
\end{document}